\documentclass[aps,pra,reprint,showpacs]{revtex4-1}

\usepackage{graphicx}
\usepackage{dcolumn}
\usepackage{bm}
\usepackage{color}
\usepackage{scrextend}		
\usepackage{url}
\usepackage{epstopdf}
\usepackage{hyperref}

\begin{document}

\title{Coupled skyrmion breathing modes in synthetic ferri- and antiferromagnets}

\author{Martin Lonsky}
\email{Electronic mail: lonsky@illinois.edu}
\author{Axel Hoffmann}
\affiliation{Materials Research Laboratory and Department of Materials Science and Engineering, University of Illinois at Urbana-Champaign, Urbana, Illinois 61801, USA}

\date{\today}

\begin{abstract}
We present micromagnetic simulations of the dynamic GHz-range resonance modes of skyrmions excited by either out-of-plane ac magnetic fields or spin torques in prototypical synthetic ferri- and antiferromagnetic trilayer structures. The observed features in the calculated power spectra exhibit a systematic dependence on the coupling strength between the individual magnetic layers and are related to pure in-phase and anti-phase breathing modes as well as to hybridizations of breathing and spin-wave modes that are characteristic for the considered circular-shaped geometry. The experimental detection of these resonant oscillation modes may provide a means for skyrmion sensing applications and for the general characterization of skyrmion states in multilayer stacks with antiferromagnetic interlayer exchange coupling.                    
\end{abstract}


\maketitle

\section{Introduction} \label{intro}
Magnetic multilayers that combine strong spin-orbit interaction with broken inversion symmetry can give rise to the presence of topologically nontrivial spin textures, so-called magnetic skyrmions \cite{Roesler2006, Muhlbauer2009}, at room temperature \cite{Jiang2017}. Several studies have indicated strongly enhanced propagation velocities of skyrmions in antiferromagnets (AFMs) and compensated ferrimagnets \cite{Caretta2018, Zhang2016, Woo2018, Barker2016}. Furthermore, only recently has the stabilization of antiferromagnetic skyrmions in synthetic AFMs been demonstrated experimentally \cite{Legrand2019, Dohi2019, Chen2020}.   
Despite the widely acknowledged superiority of such systems over ferromagnets---mainly owing to the fast current-driven motion of skyrmions due to the suppression of the skyrmion Hall effect, and the possible stabilization of very small-sized skyrmions---there still exist a number of challenges with regard to the realization of technological applications based on this particular class of multilayer structures. For instance, a straightforward electrical detection of skyrmions in all types of materials is highly challenging due to the small size of the measurement signal \cite{Wang2019}, while magnetic sensing of these spin textures in systems with vanishing magnetization is naturally impractical. 
A promising approach towards the detection and detailed characterization of skyrmion states in (synthetic) antiferromagnetic and compensated ferrimagnetic multilayers is given by harnessing the intrinsic resonances of skyrmions, such as breathing modes, which entail an oscillation of the skyrmion size at characteristic GHz frequencies \cite{Mochizuki2012, Onose2012, Garst2017, Lin2014, Schuette2014}. More specifically, in analogy to previous works on the spectral analysis of topological defects in artificial spin-ice lattices \cite{Gliga2013}, the application of broadband microwave impedance spectroscopy may offer a direct means for skyrmion sensing, that is, ascertaining the presence or absence of skyrmions and possibly even quantifying skyrmion densities. In addition to that, measurements of magnetoresistive or anomalous Hall effect signals modulated by the periodic oscillation of the skyrmion size constitute---in analogy to previous work on magnetic vortices \cite{Lendinez2020, Cui2015}---a promising approach towards electrical skyrmion detection with a high signal-to-noise ratio.      
However, in order to reliably exploit the breathing modes for electrical detection or other applications, the effect of magnetic compensation on these dynamic excitations needs to be clarified. 
As will be discussed in the present work, the excitation frequencies and magnitudes of breathing modes are highly sensitive to the strength of various competing magnetic interactions prevailing in magnetic multilayers, and thus the experimental detection of these resonant excitations may help to determine characteristic materials parameters such as exchange interactions or anisotropies. In general, besides their utility for skyrmion sensing, breathing modes were also discussed to be exploited as information carriers in data processing devices \cite{Xing2020, Lin2019, Kim2018, Seki2020}. Furthermore, it was demonstrated that these dynamic excitations are also highly relevant for skyrmion-based magnonic crystals \cite{Zhang2019, Ma2015}. One example is given by the strong coupling of breathing modes in one-dimensional skyrmion lattices in ferromagnetic thin-film nanostrips, where the existence of in-phase and anti-phase modes was demonstrated \cite{Kim2018}. In detail, the propagation of the breathing modes through the nanostrips was shown to be controllable by the strength of the applied magnetic field.       

The present work is devoted to skyrmion breathing dynamics in synthetic antiferromagnets composed of ultrahin layers that exhibit a circular-shaped geometry. Previous theoretical studies have demonstrated that the dynamic excitation modes of skyrmions are strongly influenced by the geometry of the considered system. For instance, in the case of an ultrathin circular ferromagnetic dot with a single skyrmion, the breathing modes hybridize with the radial spin-wave eigenmodes of the dot \cite{Kim2014, Mruczkiewicz2017}.  
For the case of synthetic AFMs, so far only the gyration modes of skyrmions have been studied by means of micromagnetic simulations \cite{Xing2018}. In this case, the application of time-varying in-plane magnetic fields and the presence of antiferromagnetic coupling can lead to clockwise (CW) and counterclockwise (CCW) rotation modes as well as coupled excitation modes (CW-CW, CCW-CCW, and CW-CCW).    
However, breathing modes in synthetic AFMs, which are excited by the application of out-of-plane ac magnetic fields, have not been investigated yet. Even though in Ref.\ \cite{Kravchuk2019} the spin eigenexcitations of a skyrmion in a collinear uniaxial antiferromagnetic thin film were investigated by means of numerical and analytical methods, the dynamic behavior is expected to be different in synthetic AFMs, where the interlayer exchange coupling is much weaker than the direct exchange in crystalline AFMs. In other words, there is a stronger separation of the two magnetic subsystems in synthetic AFMs, which also implies the presence of small dipolar fields \cite{Legrand2019, Duine2018}. More generally, synthetic AFMs can be viewed as materials with properties in between those of AFMs and ferromagnets \cite{Duine2018}.
Lastly, we note that the magnetization dynamics in synthetic AFMs typically exhibits a higher complexity than in ferromagnets. For instance, several studies have demonstrated the existence of resonant optic modes in synthetic AFMs in addition to the conventional acoustic (Kittel) mode in ferromagnets \cite{Waring2020, Khodadadi2017, Sorokin2020}. Due to the complicated dynamics reported for regular synthetic antiferromagnets which do not host skyrmions, it can be expected that the skyrmion breathing modes in such systems will be significantly altered compared to those in ferromagnets.   

Building on the numerous existing studies, in the present work we examine the characteristic skyrmion breathing modes in antiferromagnetic-exchange coupled disks using micromagnetic simulations with the aim of providing guidance for the experimental detection and practical application of these excitations.                  

\section{Simulation Model and Methods}
\begin{figure}
\centering
\includegraphics[width=8.5 cm]{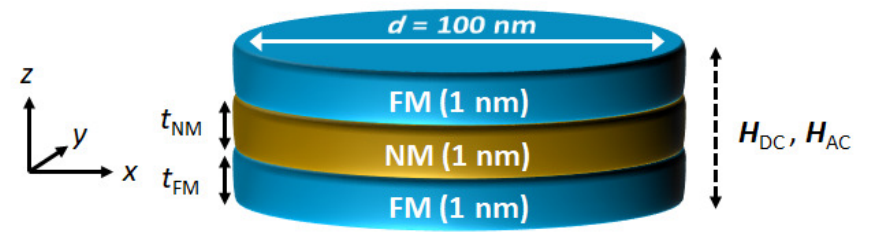}
\caption{Synthetic AFM structure consisting of two ferromagnetic (FM) and one nonmagnetic (NM) circular-shaped layers with thickness $t_{\mathrm{NM}}=t_{\mathrm{FM}}=1\,$nm and diameter $d=100\,$nm for each disk. Ac and dc magnetic fields are applied along the $z$-direction perpendicular to the layer planes.} 
\label{MODEL}%
\end{figure}%
This study focuses on a synthetic antiferromagnetic structure as depicted in Fig.\ \ref{MODEL}. The top and bottom layers are ferromagnetic materials which are separated by a nonmagnetic metallic spacer layer. In analogy to Ref.\ \cite{Xing2018}, the ferromagnetic layers are coupled via the nonmagnetic spacer through a Ruderman-Kittel-Kasuya-Yosida (RKKY) interaction which is modeled by the following energy term \cite{Parkin1991, Bruno1991}:    
\begin{equation}\label{eq:RKKY}
E_{\mathrm{RKKY}}=\frac{\sigma}{t_{\mathrm{NM}}}\left(1-\textbf{m}_{\mathrm{t}}\textbf{m}_{\mathrm{b}}\right).
\end{equation}
Here, $\sigma$ denotes the surface exchange coefficient which depends on the thickness $t_{\mathrm{NM}}$ of the nonmagnetic spacer layer and is assumed to be negative throughout this work, thus implying an antiferromagnetic coupling. Furthermore, $\textbf{m}_{\mathrm{t}}$ and $\textbf{m}_{\mathrm{b}}$ are the unit vectors of magnetization for the top and bottom layer, respectively.   

The static and dynamic states of the skyrmions in the considered model system were studied by using the Object Oriented MicroMagnetic Framework (\textsc{oommf}) code \cite{Donahue1999, Rohart2013} which carries out a numerical time integration of the Landau-Lifshitz-Gilbert (LLG) equation of motion for the local magnetization:
\begin{equation} \label{eq:LLG}
\frac{\mathrm{d}\textbf{m}}{\mathrm{d}t}=-|\gamma_{0}|\textbf{m}\times \textbf{H}_{\mathrm{eff}}+\alpha \textbf{m}\times \frac{\mathrm{d}\textbf{m}}{\mathrm{d}t}.
\end{equation} 
Here, $\gamma_{0}$ denotes the gyromagnetic constant, $\alpha$ is the Gilbert damping parameter, $\textbf{m}$ is the unit vector of the magnetization, and $\textbf{H}_{\mathrm{eff}}$ is the effective magnetic field which is proportional to the derivative of the total micromagnetic energy $U$ with respect to the magnetization. In our model, we assume that $U$ includes the Zeeman energy, the isotropic exchange interaction characterized by the exchange stiffness $A$, demagnetization effects, a uniaxial anisotropy perpendicular to the layers, and an interfacial Dzyaloshinskii-Moriya interaction (DMI)
\begin{equation}
U_{\mathrm{DMI}}=D\left[m_{z}\left(\nabla \cdot \textbf{m}\right)-\left(\textbf{m} \cdot \nabla \right)m_{z} \right],
\end{equation} 
where $D$ is the DMI constant specifying the strength of the interaction \cite{Dzyaloshinskii1958, Moriya1960}.

In order to allow for direct comparability with results for the single ferromagnetic dot considered in Ref.\ \cite{Kim2014}, we utilized identical simulation parameters for the ferromagnetic layers studied in this work. Consequently, each of the two layers corresponds to a model thin-film perpendicular anisotropy system with an exchange stiffness $A=15\,$pJ/m, perpendicular anisotropy constant $K_{\mathrm{u}}=1\,$MJ/m$^3$, saturation magnetization $\mu_{0}M_{\mathrm{s}}=1\,$T (unless specified otherwise), layer thickness $t_{\mathrm{FM}}=1\,$nm, and disk diameter $d=100\,$nm. The RKKY-coupling strength through the equally sized nonmagnetic spacer layer is varied from $\sigma=0$ to $\sigma=-3\times 10^{-3}\,$J/m$^2$. The strength of the interfacial DMI is fixed as $D=3\,$mJ/m$^2$ for both ferromagnetic layers. This value is considerably higher than in the case of Ref.\ \cite{Legrand2019}, where $D=0.2$--$0.8\,$mJ/m$^2$, whereas it lies well within the parameter range ($D=2.5$--$4.5\,$mJ/m$^2$) utilized in Ref.\ \cite{Kim2014}. As demonstrated in Refs.\ \cite{Sampaio2013, Kim2014}, higher DMI energy typically leads to larger skyrmion diameters. However, due to the choice of a smaller $M_{\mathrm{s}}$ and larger $A$ than in Ref.\ \cite{Legrand2019}, we expect to achieve comparable skyrmion diameters.         
Here, a simulation mesh with $64\times 64\times 3$ finite difference cells is defined, implying a cell size of $1.5625\times 1.5625\times 1\,$nm. 
The initial magnetization state was assumed to be that of one N\'{e}el-type skyrmion in each of the two ferromagnetic layers. The ground state has been determined by relaxing the system for $5\,$ns with a high damping constant of $\alpha = 0.5$, whereby the time evolution of the magnetization was monitored to confirm that an equilibrium state has been reached. For most of the simulations, a dc magnetic field $\mu_{0}H_{\mathrm{dc}}=50\,$mT along the perpendicular $z$-direction has been applied. As will be shown further below, such a small symmetry-breaking dc field solely increases the magnitude of certain dynamic modes, but does not have any impact on the qualitative conclusions that will also be valid for $\mu_{0}H_{\mathrm{dc}}=0\,$mT.      
Subsequently, the dynamics of the skyrmions was studied for the application of a spatially uniform time-varying ac magnetic field $H_{\mathrm{ac}}=H_{0}\sin(2\pi f t)$ along the $z$-axis, where $\mu_{0}H_{0}=0.5\,$mT is the amplitude and $f=100\,$GHz the frequency. Simulations were performed for two different scenarios, which will be discussed in more detail in Sec.\ \ref{RESULTS}. In the first case, the ac field was applied across all three layers, while in the second case only one magnetic layer was exposed to the time-varying field. While the first scenario is more realistic with regard to future experimental work, the second case will prove to be instrumental for a general understanding of the skyrmion dynamics in synthetic AFMs. It will be further shown that the qualitative results for both pictures share many similarities.   
Another important point is that the obtained results are nearly identical regardless of whether the ac field is applied over the entire simulation time or for only a limited time period after which the data are recorded. 
Moreover, we note that there exist further experimental approaches to excite the GHz-range dynamic modes of skyrmions, such as, for example the application of laser or heat pulses \cite{Ogawa2015}, or spin torques. The latter will also be incorporated into the micromagnetic simulations and briefly discussed in the final part of Sec.\ \ref{RESULTS}. 
       
For the skyrmion dynamics simulations the damping parameter has been chosen as $\alpha=0.01$ to ensure a good frequency resolution of the excited modes. The dynamics is simulated for at least $5\,$ns with data taken every $2\,$ps. Eventually, the power spectral density (PSD) of the spatially-averaged $z$-component of the magnetization $\langle m_{z}\rangle (t)$ is calculated by using a fast Fourier transform (FFT) algorithm. As a proof of concept, the results from Ref.\ \cite{Kim2014} for a single ferromagnetic layer had been reproduced before numerical calculations were conducted for the synthetic AFM. In addition to the synthetic AFM trilayer, Sec.\ \ref{RESULTS} also includes a discussion of micromagnetic simulations carried out on a model system for a synthetic ferrimagnet with unbalanced antiparallel moments in the two ferromagnetic layers.                      

\section{Results and Discussion} \label{RESULTS}
\subsection{Static Properties of Skyrmions in Synthetic Antiferromagnets}
\begin{figure}
\centering
\includegraphics[width=8.5 cm]{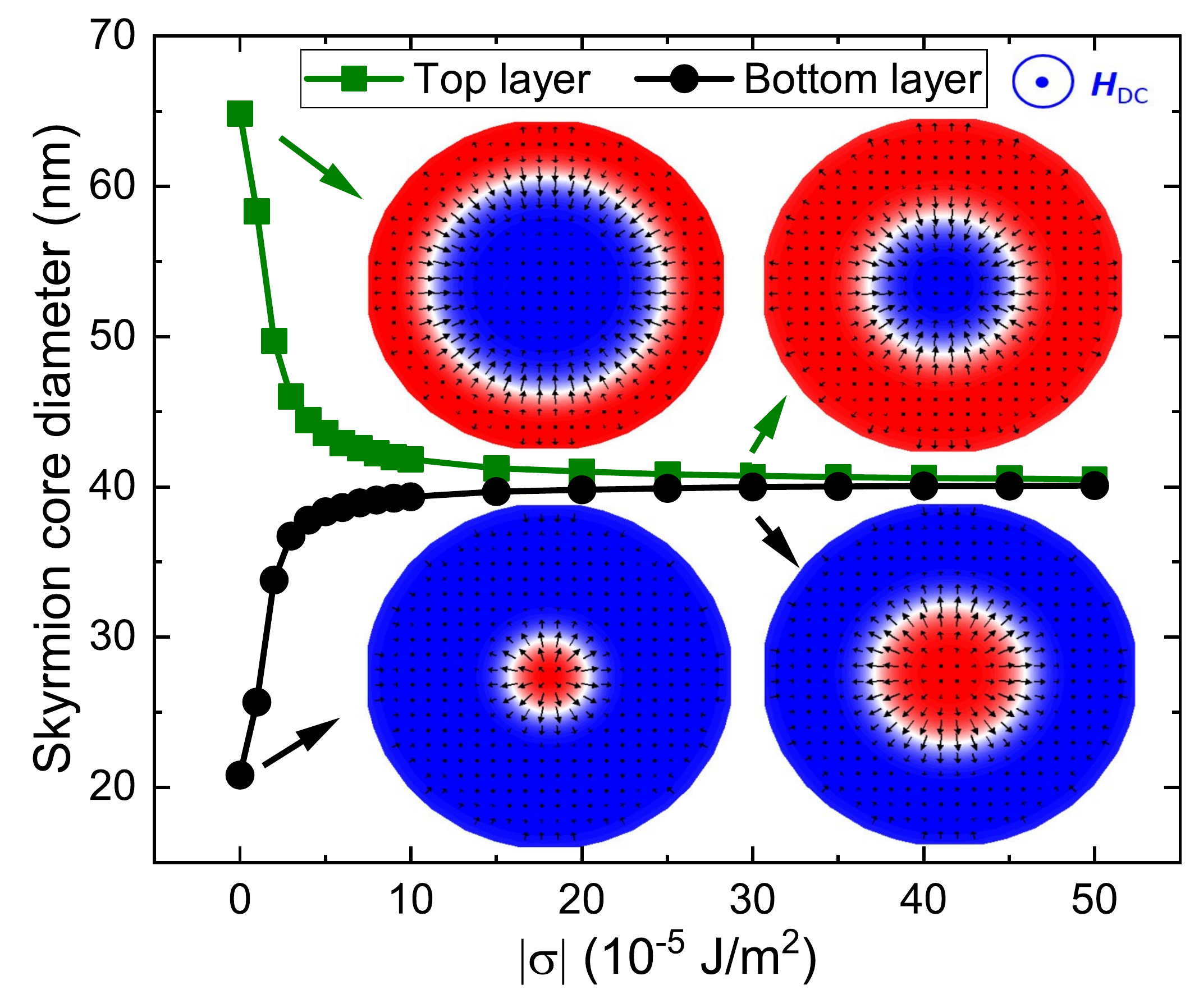}
\caption{Skyrmion core diameter plotted against the absolute value of the antiferromagnetic coupling strength $\sigma$ for the top and bottom layer. Selected skyrmion ground states are shown for $\sigma=0\,$J/m$^2$ and $\sigma=-3\times 10^{-4}\,$J/m$^2$ for the individual layers. A dc magnetic field $\mu_{0}H_{\mathrm{dc}}=50\,$mT was applied along the positive $z$-direction as shown in the top right of the diagram.} 
\label{SKSIZE}%
\end{figure}%
As a first step, the ground states were calculated for different antiferromagnetic coupling strengths at a dc magnetic field $\mu_{0}H_{\mathrm{dc}}=50\,$mT. The skyrmion core diameter, here defined as twice the distance from the center ($|m_{z}|=1$) to the point where $m_{z}=0$, is plotted against the coupling strength $|\sigma|$ in Fig.\ \ref{SKSIZE} for the top and the bottom layer. Furthermore, skyrmion states for two selected coupling strengths $\sigma=0\,$J/m$^2$ and $\sigma=-3\times 10^{-4}\,$J/m$^2$ are illustrated. As shown in the top right of the diagram, $H_{\mathrm{dc}}$ was applied along the positive $z$-direction. It can be seen that at high coupling strengths the diameters of both skyrmions are almost identical, $d_{\mathrm{top}}\approx d_{\mathrm{bottom}}\approx 40$nm, while towards lower values of $|\sigma|$ the size difference increases rapidly up to $\Delta d=d_{\mathrm{top}}-d_{\mathrm{bottom}}\approx 45$nm for $\sigma=0\,$J/m$^2$. This can be explained by the magnetostatic interaction becoming more relevant for the total micromagnetic energy in the case of weaker interlayer exchange coupling. More generally, as a result of the various competing energy terms, the skyrmion size is also highly sensitive to variations of other parameters such as the DMI strength, the exchange stiffness or the saturation magnetization (see discussion further below). Even though the DMI energy utilized in the present study is larger than in the case of Ref.\ \cite{Legrand2019}, due to the differences in other parameters, such as $A$ and $M_{\mathrm{s}}$, we obtain similar values for the skyrmion core diameter in the case of comparable coupling strengths $|\sigma|\approx 2\times 10^{-4}\,$J/m$^2$.          

\begin{figure}
\centering
\includegraphics[width=8.6 cm]{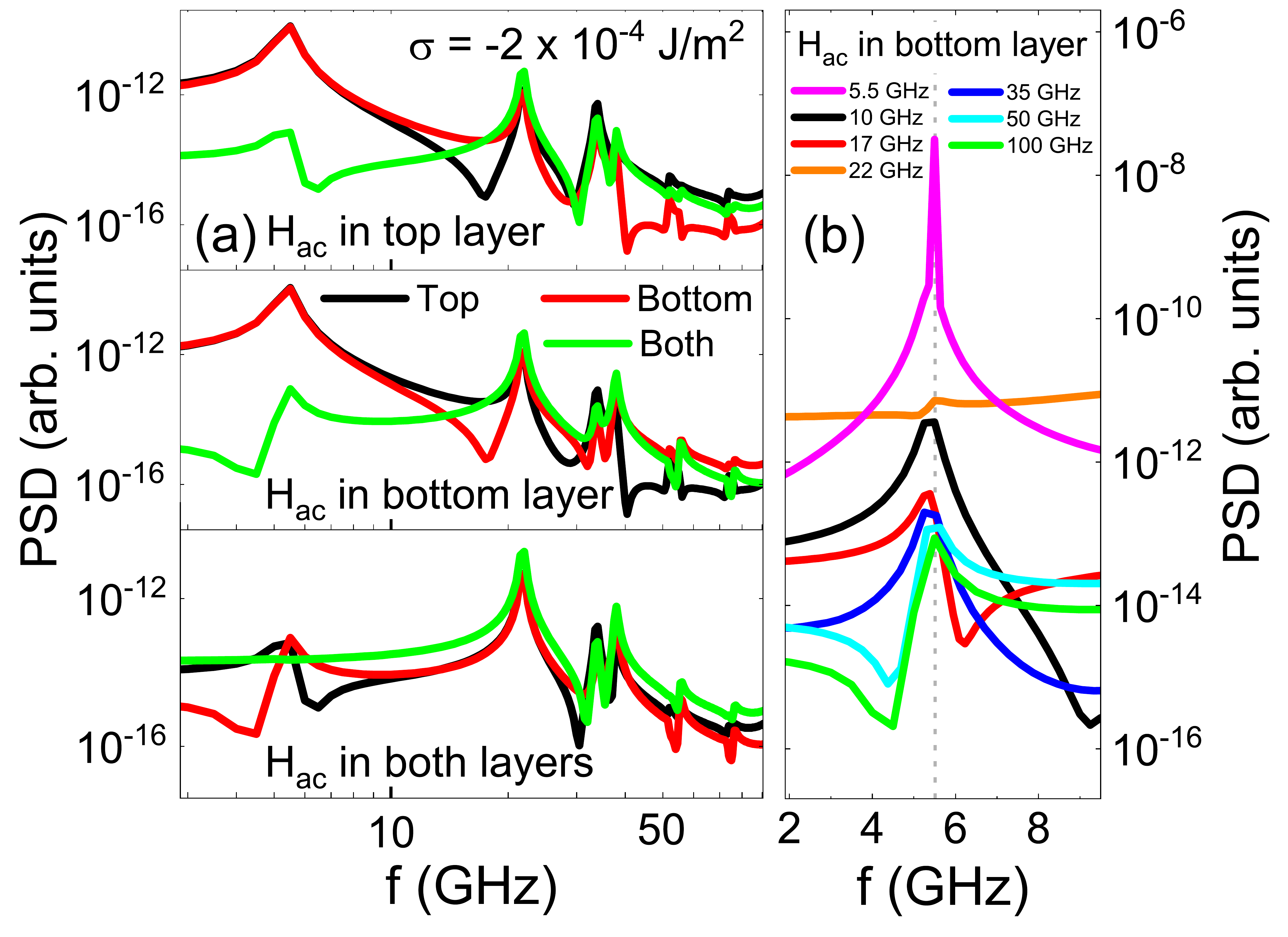}
\caption{(a) Power spectral density (PSD) of $\langle m_{z}\rangle (t)$ calculated in the top, bottom, and both layers for a selected $\sigma$, where the ac magnetic field with $f=100\,$GHz is applied either across the top layer, bottom layer or both layers. (b) Line shape of mode 1 in dependence of the excitation frequency (PSD calculated for both magnetic layers).} 
\label{DIFFLAYERS}%
\end{figure}%
\subsection{Breathing Modes in Synthetic Antiferromagnets}
An overview of the dynamic response to an applied ac magnetic field with $f=100\,$GHz and amplitude $\mu_{0}H_{0}=0.5\,$mT is given in Fig.\ \ref{DIFFLAYERS}. The calculated spatially-averaged PSD of $\langle m_{z}\rangle (t)$ for the top (black curves), bottom (red curves) and both layers (green curves) in the particular case of $\sigma=-2\times 10^{-4}\,$J/m$^2$ is shown in Fig.\ \ref{DIFFLAYERS}(a) as a function of frequency. The time-varying field is either applied across the top layer, the bottom layer or all layers. There are several distinct features at identical frequencies for all three cases with only minimal differences in the peak heights, except for the lowest-lying excitation at $f=5.5\,$GHz, where in the case of the ac field being applied across all layers the peak amplitude is strongly suppressed. This feature can only be observed in an enlarged plot, or alternatively occurs in a more pronounced way when the frequency of the ac field is reduced until it matches the resonance frequency (not shown). It will be discussed further below that sufficiently high dc magnetic fields lead to an enhanced magnitude of this lowest-lying excitation---even for the case that all layers are exposed to the time-varying magnetic field.   
We emphasize that the peak magnitude and line shape of mode 1 strongly depend on the frequency of the ac magnetic field. This can be seen in Fig.\ \ref{DIFFLAYERS}(b) for the case of the ac field applied across the bottom layer. This indicates that the interplay between the first mode and the other resonant modes, which become activated towards higher excitation frequencies, strongly affects the observed line shape due to complicated phase relationships. For instance, utilizing an excitation frequency resonant with mode 2 results in a strong suppression of mode 1 (cf.\ orange curve). Moreover, the activation of further higher-order modes can change the symmetry of the resonance peak. For some excitation frequencies, there is a pronounced antiresonance adjacent to the resonance peak.

\begin{figure}
\centering
\includegraphics[width=8.6 cm]{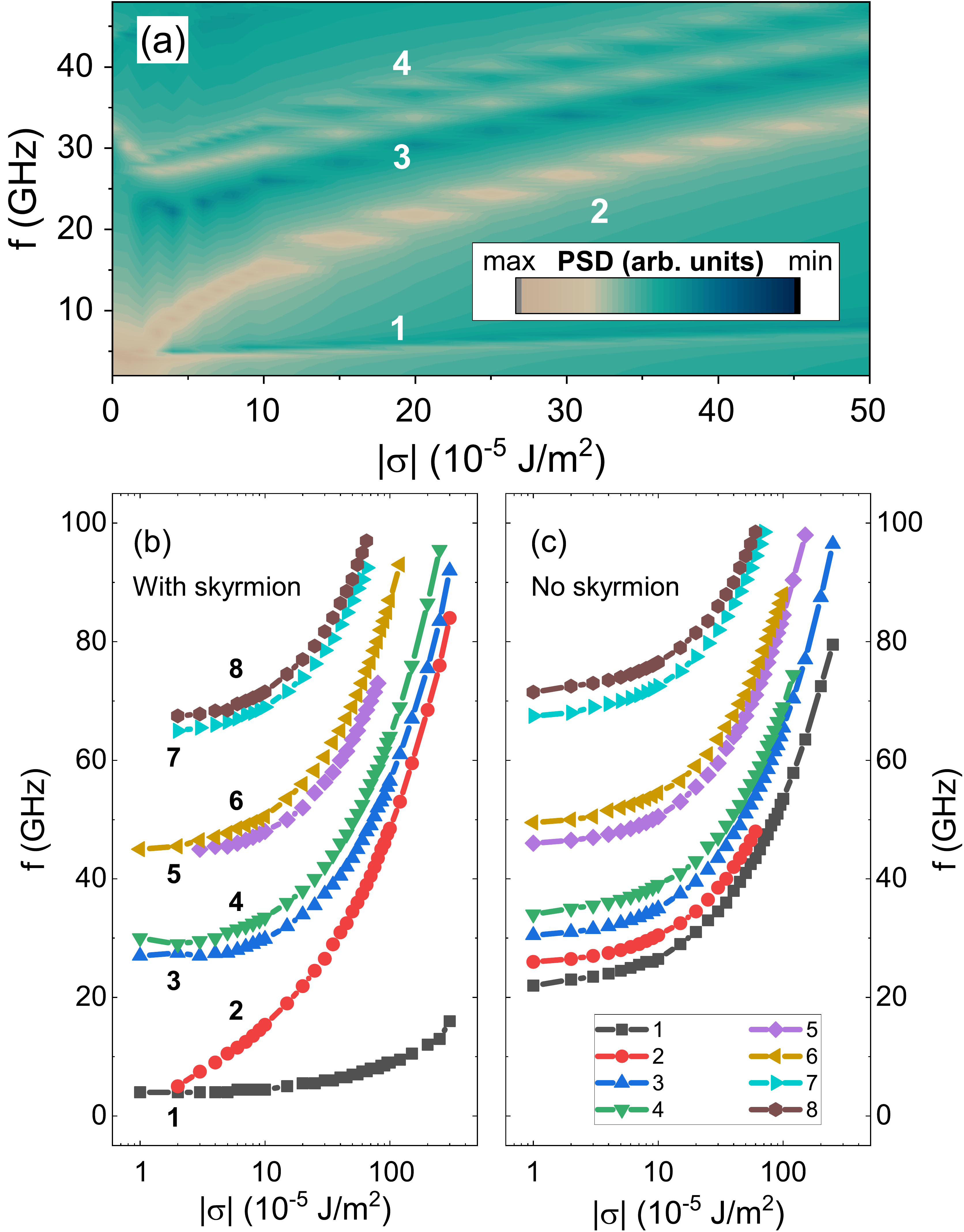}
\caption{(a) Map of the PSD as a function of the interlayer exchange coupling strength $|\sigma|$ and the frequency with an external ac magnetic field applied across the top layer. Only the four lowest-lying resonance modes are shown. (b) Resonance frequency for all eight modes in the entire range of $|\sigma|$. (c) Resonance frequency for the first eight dynamic modes in the uniform ground state of the synthetic AFM without the presence of skyrmions.} 
\label{PSDMAP}%
\end{figure}%
Figure \ref{PSDMAP}(a) shows a map of the PSD as a function of the antiferromagnetic coupling and the frequency for the scenario that the ac field is only present in the top layer. Notice that only the four lowest-lying resonance modes are depicted. Furthermore, gaps in the individual branches occur due to the discrete nature of the utilized antiferromagnetic coupling strength $\sigma$. 
It is evident that the characteristic frequencies of the resonances and antiresonances shift towards higher values upon increasing the interlayer coupling. This is further clarified in Fig.\ \ref{PSDMAP}(b) for all eight resonance modes and the complete range of the antiferromagnetic coupling strength $\sigma$. From this logarithmic representation it is clear that mode 2 displays the strongest increase in dependence of $\sigma$. Furthermore, as will be proven by the discussion further below, the observed modes occur in pairs of interrelated resonances. Lastly, Fig.\ \ref{PSDMAP}(c) displays the resonance frequencies for eight modes in the case of the uniform ground state of the considered synthetic AFM, that is, where no skyrmion is present. The two lowest-lying modes only occur in this magnetization state and exhibit an entirely different dependence on $\sigma$ in comparison to modes $1$ and $2$ in the skyrmion state. In detail, these resonances can be explained by the dynamics at the edges of the circular-shaped synthetic AFM structure, where the magnetization is tilted as a result of the boundary conditions related to the DMI, see Ref.\ \cite{Kim2014}. Such edge modes are also the physical origin of all higher-order resonances. Interestingly, their dependence on the interlayer coupling strength is almost identical to modes $3$--$8$ in Fig.\ \ref{PSDMAP}(b). In fact, as will be shown in the following, these edge modes are found to hybridize with the characteristic skyrmion eigenexcitations.          

\subsubsection{Visualization of Dynamic Modes}
\begin{figure}
\centering
\includegraphics[width=8.6 cm]{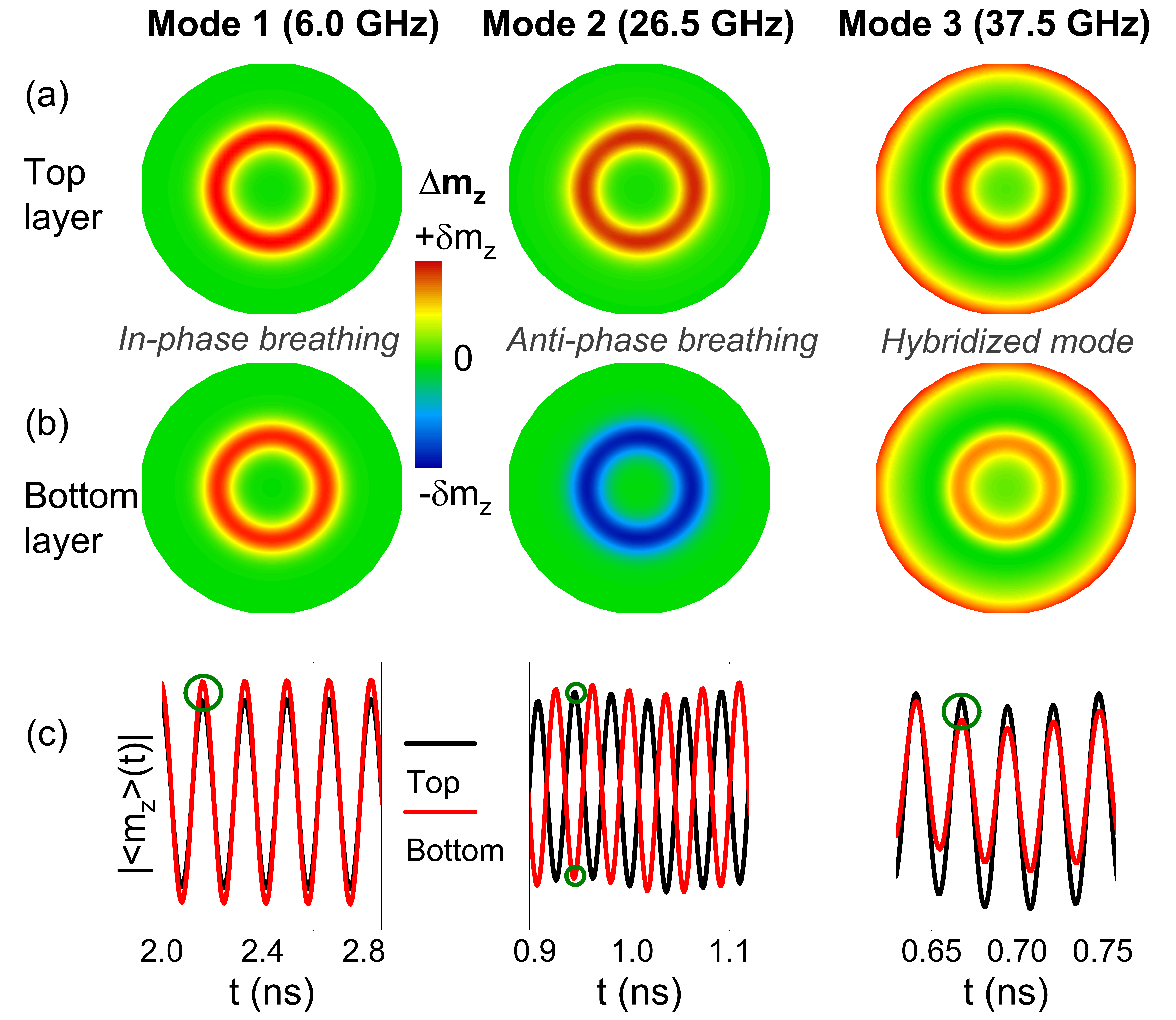}
\caption{Snapshots of resonance modes 1 to 3 shown in three columns for $\sigma=-3\times 10^{-4}\,$J/m$^2$ and a time-varying magnetic field with the respective excitation frequency applied across the top layer. Two-dimensional contour plots display the difference $\Delta m_{z}$ in the magnetization component $m_{z}$ between the ground state at $t=0$ and an excited state with maximal amplitude $\langle m_{z}\rangle(t_{\mathrm{m}})$ at a selected time $t_{\mathrm{m}}$ for (a) the top and (b) the bottom layer. (c) Absolute value of the time-dependent spatially-averaged magnetization, $|\langle m_{z}\rangle(t)|$, for top and bottom layers with arbitrary scaling of the y-axes. Green circles indicate the selected $t_{\mathrm{m}}$ for the snapshots.} 
\label{SNAPSHOT}%
\end{figure}%
Hereinafter, the physical origin of the individual modes will be discussed in detail. For this purpose, we consider two-dimensional snapshots of the resonance modes for the particular case of $\sigma=-3\times 10^{-4}\,$J/m$^2$ and an ac magnetic field with the respective mode's excitation frequency, applied only across the topmost layer. Figures \ref{SNAPSHOT}(a) and (b) include contour plots illustrating the difference $\Delta m_{z}$ in the magnetization component $m_{z}$ between the ground state at $t=0$ and an excited state with maximal amplitude $\langle m_{z}\rangle(t_{\mathrm{m}})$ at a selected time $t_{\mathrm{m}}$ for modes 1 to 3 in the top and bottom layer, respectively. For the sake of better comparability of the dynamic excitations in the two antiferromagnetically coupled layers, $m_{z}$ in the bottom layer has been multiplied by a factor of $-1$ at each position. The maximum change in $m_{z}$ is denoted by $+\delta m_{z}$ (red color) and $-\delta m_{z}$ (blue color).  
In Fig.\ \ref{SNAPSHOT}(c), the time-dependent spatially-averaged $z$-component of the magnetization, $|\langle m_{z}\rangle(t)|$, for each of the two magnetic layers is shown. Green circles indicate the selected $t_{\mathrm{m}}$ for the snapshots. In the case of mode 1 ($f=6.0\,$GHz), the $m_z$ component in a ring-shaped area increases for both the top and bottom ferromagnetic disks, corresponding to a simultaneously increasing skyrmion diameter in both layers. Therefore, mode 1 corresponds to a synchronized (in-phase) breathing motion of the two skyrmions which originates in the magnetic coupling between the individual layers. In contrast to that, as indicated by the snapshots in the second column of Fig.\ \ref{SNAPSHOT}, mode 2 ($f=26.5\,$GHz) involves an anti-phase skyrmion core oscillation. While $m_{z}$ increases within the ring-shaped area in the top layer and thereby implies a larger skyrmion core diameter than in the ground state, $m_{z}$ in the bottom layer decreases, corresponding to a reduced diameter. For the case of $\sigma=-3\times 10^{-4}\,$J/m$^2$, the individual skyrmion core diameters oscillate between $39.2\,$nm and $42.0\,$nm. As a simplified classical analog, the coupled skyrmion breathing motions may be viewed as two harmonic oscillators of identical mass (e.g., pendula) which are coupled (e.g., by a spring with spring constant $k$) and subject to an external periodic driving force---cf.\ the coupled gyration modes of magnetic vortices \cite{Vogel2011, Buchanan2005}. Thus, the in-phase and anti-phase oscillations are two normal modes of the system. By regarding the antiferromagnetic coupling strength $\sigma$ as the analog of the classical spring constant $k$, it becomes immediately clear that the increasing frequency splitting between the in-phase and anti-phase breathing modes towards higher values of $|\sigma|$ (cf.\ Fig.\ \ref{PSDMAP}) is fully consistent with the classical picture where the splitting is proportional to $k$. Lastly, this classical model can also explain the presence of antiresonances in the power spectra---see, for instance, Fig.\ \ref{DIFFLAYERS}(b), where this feature is most pronounced at higher excitation frequencies---as this is a well-known phenomenon in the physics of coupled oscillators.           
        
The above-described behavior strongly resembles the results in Ref.\ \cite{Kim2018} for micromagnetic simulations of coupled breathing modes in a one-dimensional skyrmion lattice in thin-film nanostrips.      
However, a striking difference is that in the case of the synthetic AFM studied in the present work, the in-phase breathing motion exhibits a lower energy than the anti-phase oscillation, while the reverse is true for coupled skyrmions in a nanostrip. As presented in Fig.\ \ref{PSDMAP}(b), the relationship between the coupling strength $\sigma$ and the resonance frequency is different for the in-phase and anti-phase breathing modes, and their energy difference can be controlled by varying the antiferromagnetic coupling parameter, which, for instance, in practice is related to the thickness of the spacer layer. Towards lower absolute values of $\sigma$, the energy splitting between the two modes becomes increasingly smaller up to the point where for $|\sigma| \leq 1\times 10^{-5}\,$J/m$^2$ only the in-phase mode can be identified unambiguously. Due to the reduced (or even vanishing) interlayer exchange coupling, the breathing mode amplitude in the bottom layer is observed to be considerably smaller than in the top layer which has been excited by the ac magnetic field. Consequently, the anti-phase mode is not detectable in this case.     
In analogy to the work on thin-film nanostrips \cite{Kim2018}, the frequency splitting for the antiferromagnetically-coupled skyrmions can also be explained by a symmetry breaking of the potential energy profile compared to the case of an isolated skyrmion as studied in Ref.\ \cite{Kim2014}. In fact, a similar behavior was also reported for coupled gyration modes of magnetic vortices \cite{Jung2011, Lee2011, Han2013} and skyrmions \cite{Kim2017}. In all these cases, the in-phase motion exhibits a lower energy than the anti-phase mode. The fact that this is also true for breathing modes in a synthetic AFM but the opposite effect is observed for dipolar-coupled breathing modes in thin-film nanostrips \cite{Kim2018} implies that the frequency splitting is highly sensitive to the interplay of various different magnetic interactions.

In addition to the pure breathing modes and similar to the case of a single skyrmion in an ultrathin ferromagnetic dot \cite{Kim2014}, the higher-order modes in Fig.\ \ref{PSDMAP} correspond to the hybridization of the breathing motion with geometrically quantized spin wave eigenmodes of the individual circular-shaped layers. Exemplary snapshots for mode 3 ($f=37.5\,$GHz) in Fig.\ \ref{SNAPSHOT} illustrate such a hybridization for the case of the synthetic AFM. In analogy to the pure breathing modes, each of the higher-order hybridization modes also occurs in an in-phase and anti-phase variation with different energies.

Lastly, we note that for the extended case of several antiferromagnetically-coupled pairs of layers, additional peaks emerge in the calculated power spectra. These new resonances are related to further coupled breathing modes with different phase shifts between the individual skyrmion eigenexcitations---similar to the dynamics of a one-dimensional skyrmion lattice in a ferromagnetic nanostrip \cite{Kim2018}. Such synthetic AFMs with an increased number of magnetic layers have been discussed to host antiferromagnetic skyrmions with enhanced thermal stability \cite{Legrand2019}. Consequently, the spectral analysis of the inherent dynamic eigenexcitations in these extended multilayer systems may also be practically relevant. However, a detailed discussion of this significantly more complex scenario is beyond the scope of the present work. Instead, the remainder of this paper contains a more in-depth analysis of skyrmion breathing modes in a system with only two antiferromagnetically-coupled layers.     

\subsubsection{Role of External Magnetic Field $H_{\mathrm{dc}}$}
\begin{figure}
\centering
\includegraphics[width=8.6 cm]{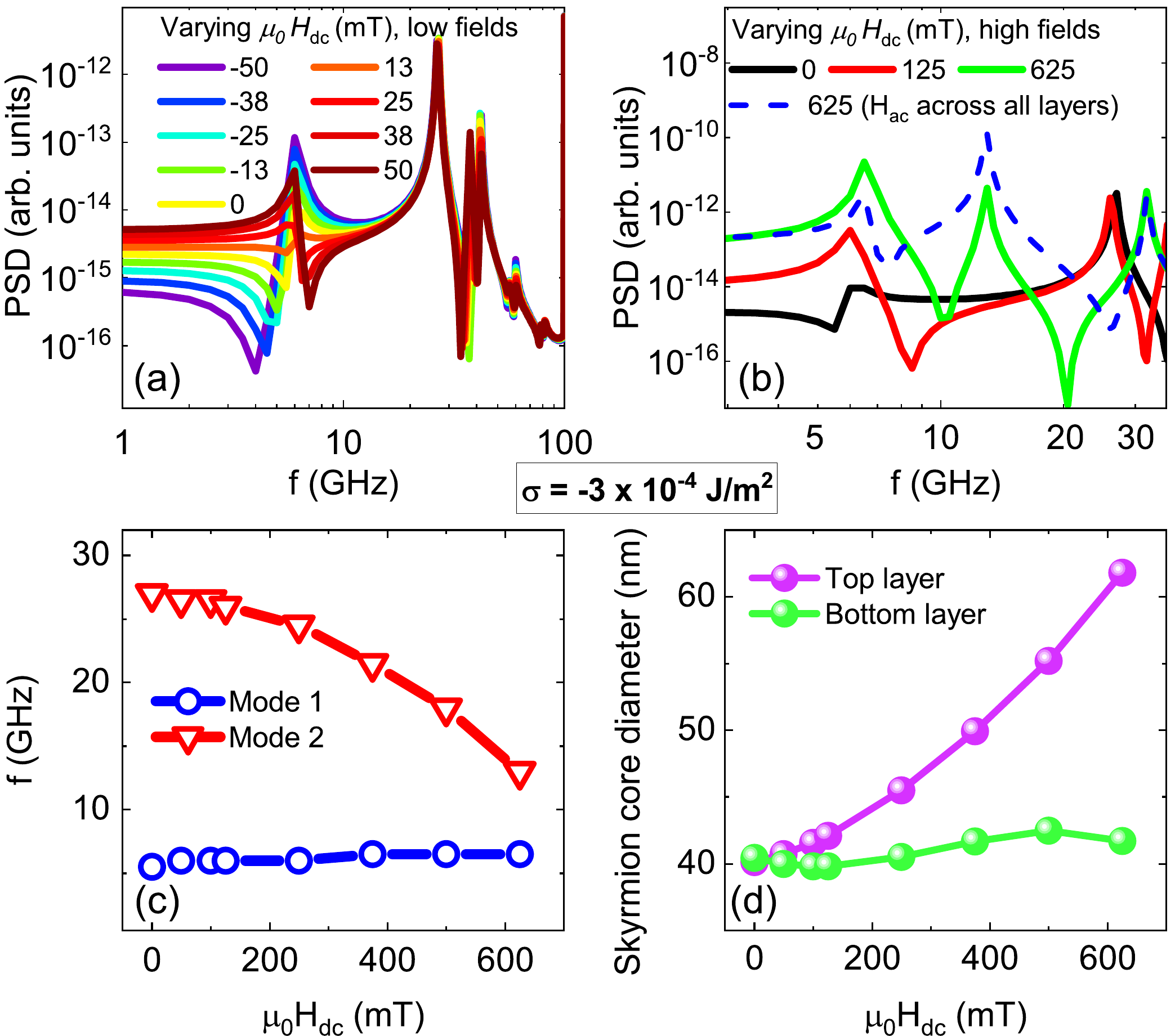}
\caption{Dependence on external dc magnetic field $H_{\mathrm{dc}}$ ($H_{\mathrm{ac}}$ applied across top layer only) for $\sigma=-3\times 10^{-4}\,$J/m$^2$. (a) Power spectra for small values of $H_{\mathrm{dc}}$. Amplitude and line shape of mode 1 change systematically. (b) PSD for selected higher values of $H_{\mathrm{dc}}$ with strong changes of modes 1 and 2. Dashed line shows a spectrum for large $H_{\mathrm{dc}}$ with $H_{\mathrm{ac}}$ applied across all layers. (c) Resonance frequency for mode 1 increases slowly as a function of $H_{\mathrm{dc}}$, while a strong decrease is observed for mode 2. (d) Skyrmion core diameter for both ferromagnetic layers in dependence of $H_{\mathrm{dc}}$.} 
\label{HDCDEP}%
\end{figure}%
Figure \ref{HDCDEP} illustrates the dependence of the skyrmion breathing dynamics on the external magnetic dc field $H_{\mathrm{dc}}$ for $\sigma=-3\times 10^{-4}\,$J/m$^2$. While the dc field is present in all layers, the time-varying magnetic field is only applied to the top layer. As expected for a synthetic AFM, and in stark contrast to ferromagnets \cite{Kim2014}, applied dc fields of comparably low magnitude do not lead to observable variations of the resonance frequencies, see Fig.\ \ref{HDCDEP}(a). However, for the in-phase breathing mode at $6.0\,$GHz the peak magnitude and the line shape do depend on the dc field. First, the peak height increases with the absolute value of the dc magnetic field. In addition to that, the line shape clearly changes its symmetry at smaller positive field values between $13$ and $25\,$mT. By contrast, the higher-order modes remain nearly unaltered. Only at higher fields does the anti-phase mode shift towards lower frequencies. Two selected power spectra at higher dc fields are depicted in Fig.\ \ref{HDCDEP}(b) and compared to the simulated curve for zero field. In addition, a spectrum for $\mu_{0}H_{\mathrm{dc}}=625\,$mT is displayed for the case of the ac field being applied across all layers (dashed blue curve). Clearly, in contrast to small dc fields (cf.\ bottom panel of Fig.\ \ref{DIFFLAYERS}(a) where $\mu_{0}H_{\mathrm{dc}}=50\,$mT), the in-phase mode can be observed for this scenario due to the symmetry breaking caused by the strong dc field. Consequently, the application of a sufficiently large dc magnetic field is expected to facilitate the experimental detection of the in-phase breathing mode in a synthetic AFM. As shown in Fig.\ \ref{HDCDEP}(c), the resonance frequency of this mode remains nearly unaffected even by large dc fields, whereas for the anti-phase resonance a strong decrease can be observed. The corresponding static skyrmion core diameters in both magnetic layers are presented in Fig.\ \ref{HDCDEP}(d). While the diameter of the skyrmion located in the top layer increases significantly from about $40\,$nm to more than $60\,$nm throughout the simulated field range, the skyrmion in the bottom layer does not exhibit major changes in its spatial extent. Typically, larger external dc field values would cause the skyrmion in the bottom layer to shrink in size, but this effect is counteracted by the strong antiferromagnetic interlayer exchange coupling to the skyrmion in the top layer which would make a greater skyrmion diameter more favorable. Therefore, the green curve in Fig.\ \ref{HDCDEP}(d) represents a compromise between these two competing effects. Note that even larger values of $H_{\mathrm{dc}}$ lead to a breakdown of the skyrmion state for the given synthetic AFM. Moreover, negative values for $H_{\mathrm{dc}}$ lead to larger skyrmion diameters in the bottom layer, while the size of the skyrmion in the top layer does not undergo strong changes. The systematics for the power spectra remains the same as for positive magnetic fields. Finally, we note that the effect of an increasing difference in skyrmion core diameters towards larger $H_{\mathrm{dc}}$ values becomes less pronounced at higher interlayer exchange coupling strengths.         

\subsubsection{Dependence on Saturation Magnetization $M_{\mathrm{s}}$}
\begin{figure}
\centering
\includegraphics[width=8.6 cm]{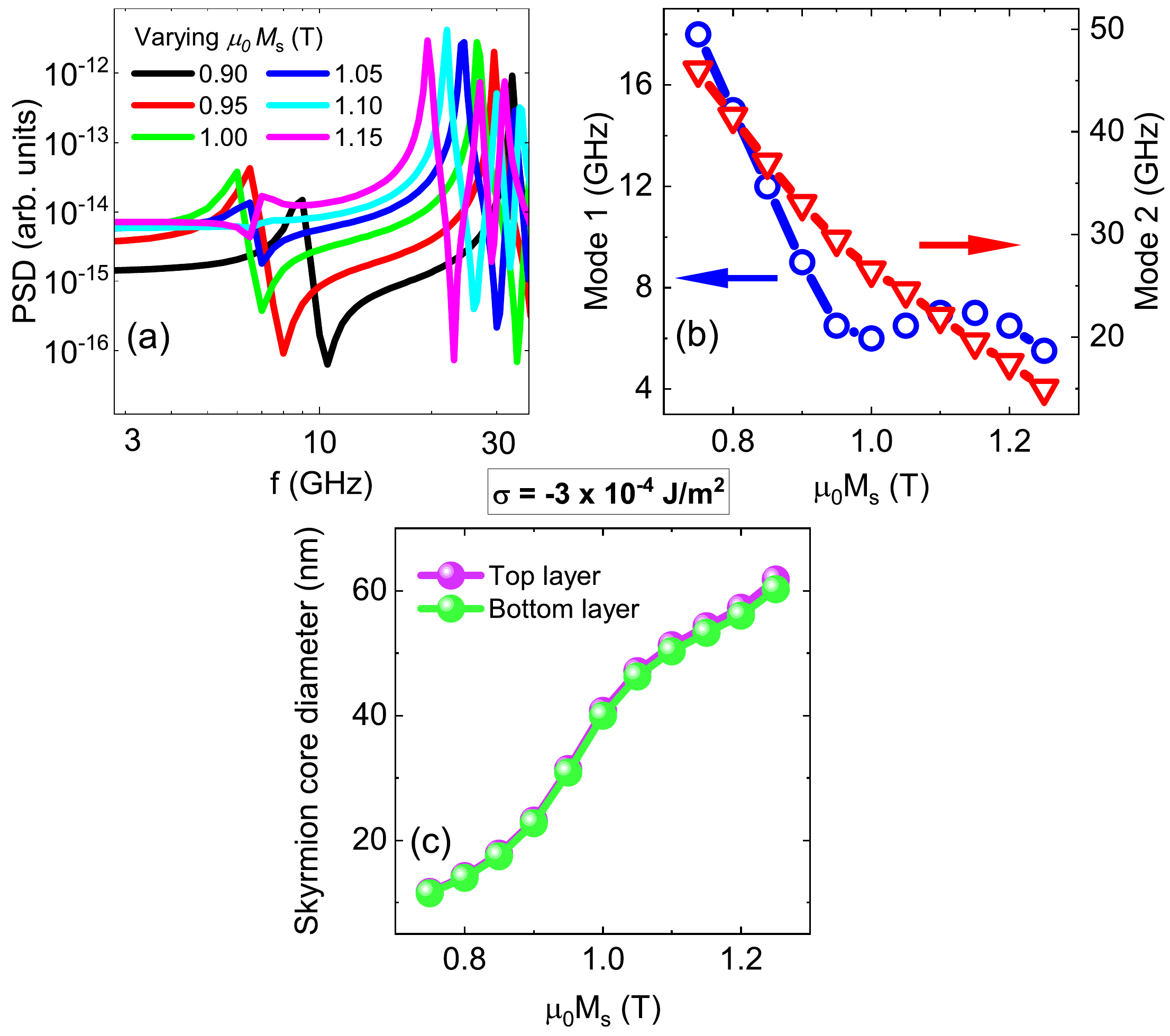}
\caption{Dependence on saturation magnetization $M_{\mathrm{s}}$ in both layers ($H_{\mathrm{ac}}$ applied across top layer only) for $\sigma=-3\times 10^{-4}\,$J/m$^2$. (a) Power spectra for selected saturation magnetization $(M_{\mathrm{s}})$ values of the ferromagnetic layers. (b) Resonance frequencies shift in a different way for modes 1 and 2 as a function of $M_{\mathrm{s}}$. (c) Skyrmion core diameter for both ferromagnetic layers in dependence of $M_{\mathrm{s}}$.} 
\label{MSANDHDEP}%
\end{figure}%
Aside from the previously discussed variations of the dc magnetic field, it is clarified in Fig.\ \ref{MSANDHDEP} that also changes in the saturation magnetization $M_{\mathrm{s}}$ of both layers can alter the skyrmion breathing dynamics in the considered synthetic AFM. In analogy to the case of small applied dc fields [cf.\ panel (a) in Fig.\ \ref{HDCDEP}], a change in the line shape symmetry of mode 1 can also be observed for variations of the saturation magnetization $M_{\mathrm{s}}$ in both magnetic layers as shown in Fig.\ \ref{MSANDHDEP}(a). This behavior also resembles the line-shape changes presented in Fig.\ \ref{DIFFLAYERS}(b) which arise due to the varying ac magnetic field frequency and can be attributed to the different phase relationships among the activated modes.   
Figure \ref{MSANDHDEP}(b) clarifies that, in addition to the line shape changes, the resonance frequency of mode 1 shifts in a nontrivial way as a function of $\mu_{0}M_{\mathrm{s}}$, while it exhibits a nearly linear behavior in the case of mode 2. At low values of $\mu_{0}M_{\mathrm{s}}$, the resonance frequency of mode 1 decreases upon increasing $\mu_{0}M_{\mathrm{s}}$ and displays a local minimum at $1.0\,$T. This is followed by a slow increase towards larger values of $\mu_{0}M_{\mathrm{s}}$, a local maximum at $1.15\,$T and a subsequent decrease. As shown in \ref{MSANDHDEP}(c), the increase in the saturation magnetization from $0.75$ to $1.25\,$T entails a growing skyrmion core diameter from around $11\,$nm up to $60\,$nm in both magnetic layers, while the relative size difference remains equally small due to the strong antiferromagnetic coupling. In analogy to Ref.\ \cite{Kim2014}, the limiting factor for the skyrmion diameter is given by the interaction with the tilted magnetization at the boundary of the nanodisks.   
In the previously discussed classical picture larger values of $M_{\mathrm{s}}$ would correspond to an increasing mass of each of the two harmonic oscillators, leading to lower eigenfrequencies. While this simple model correctly explains the behavior of mode 2, in the case of mode 1 it is only applicable for low $\mu_{0}M_{\mathrm{s}}$ values up to $1.0\,$T. For higher values of $\mu_{0}M_{\mathrm{s}}$, however, the interplay with the higher-order modes and the competition of various micromagnetic energies are the cause of an unexpected behavior. In fact, a similar systematics is observed for variations of the DMI parameter $D$ or the exchange stiffness $A$, as well as for other values of the interlayer exchange coupling strength $\sigma$.

For the experimental detection of breathing oscillations in antiferromagnetically-coupled multilayers, as well as for skyrmion sensing in general, both the in-phase and anti-phase modes are suitable candidates when the system is excited at their respective resonance frequencies. In practice, the scenario of an ac field being applied across the entire synthetic AFM is clearly more realistic than to assume its presence only within a single layer. We emphasize that the former case will require a sufficiently large dc magnetic field to break the symmetry and thus enable the detection of the in-phase breathing mode, see Fig.\ \ref{HDCDEP}(b). By contrast, the anti-phase mode is expected to be experimentally detectable in a more straightforward way. Furthermore, we point out that the unique dependence of the eigenfrequencies on the external dc field and the saturation magnetization will allow to draw detailed conclusions about the skyrmion states from the spectral analysis.      
As has been shown, it may be also worthwhile to utilize higher excitation frequencies ($f\approx 100\,$GHz) and deduce further information about fundamental magnetic parameters from the interplay of various spin excitation modes. For instance, while the magnitude of the anti-phase mode remains large for different values of $M_{\mathrm{s}}$, both the line shape and the magnitude of the in-phase mode are clearly more sensitive to variations of this magnetic parameter. 

\subsection{Breathing Modes in Synthetic Ferrimagnets}
\begin{figure}
\centering
\includegraphics[width=8.6 cm]{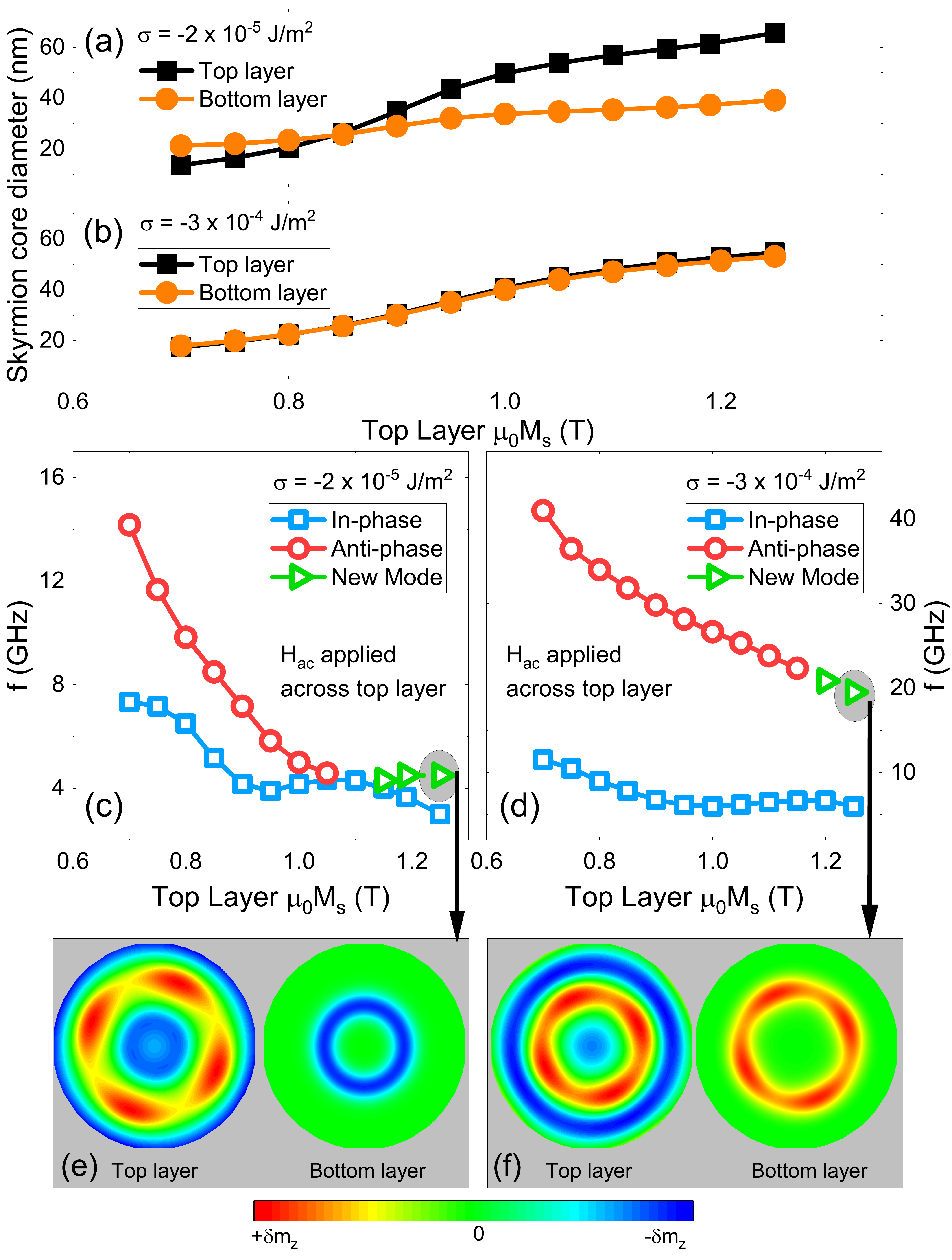}
\caption{Static skyrmion core diameters for varying $M_{\mathrm{s}}$ in the top layer and constant $\mu_{0}M_{\mathrm{s}}=1.0\,$T in the bottom layer are presented in the case of (a) $\sigma=-2\times 10^{-5}\,$J/m$^2$ and (b) $\sigma=-3\times 10^{-4}\,$J/m$^2$. Panels (c) and (d) show the resonance frequency for the two lowest-lying modes as a function of $M_{\mathrm{s}}$ in the top layer for both coupling strengths. Snapshots (calculated in analogy to Fig.\ \ref{SNAPSHOT}) of a newly emerging, non-radial dynamic mode at higher $M_{\mathrm{s}}$ are displayed in (e) and (f).} 
\label{FERRIMAG}%
\end{figure}%
In the following, we will discuss how the coupled skyrmion breathing dynamics is altered in the case of a synthetic ferrimagnet, \textit{i.e.}, the same trilayer system as depicted in Fig.\ \ref{MODEL}, but now containing unbalanced antiparallel moments in the two ferromagnetic layers. For the sake of simplicity, we assume varying values of the saturation magnetization $M_{\mathrm{s}}$ only in the top layer while keeping the bottom-layer magnetization constant at $\mu_{0}M_{\mathrm{s}}=1.0\,$T. The calculated static skyrmion core diameters for the two magnetic layers are depicted in Fig.\ \ref{FERRIMAG}(a) and (b) for two different coupling strengths $\sigma$. While for the stronger interlayer exchange coupling the skyrmion size is nearly identical in both layers, in the case of a weak coupling the individual skyrmion core diameters differ significantly over a broad range of top-layer $M_{\mathrm{s}}$ values. 
   
Fig.\ \ref{FERRIMAG}(c) and (d) displays the evolution of the resonance frequency for the two lowest-lying eigenmodes with varying top-layer $M_{\mathrm{s}}$ in the case of weak and strong antiferromagnetic coupling, respectively. Both the in-phase and anti-phase mode resonance frequencies decrease monotonically as a function of the top layer $M_{\mathrm{s}}$. While the separation between the two modes remains large throughout the entire range of $M_{\mathrm{s}}$ values in the case of strong interlayer exchange coupling ($\sigma=-3\times 10^{-4}\,$J/m$^2$), for $\sigma=-2\times 10^{-5}\,$J/m$^2$ the anti-phase resonance mode closely approaches the in-phase resonance frequency until it vanishes at $\mu_{0}M_{\mathrm{s}}=1.1\,$T, where only the in-phase mode can be observed. Towards even higher values of $M_{\mathrm{s}}$, a second resonance mode reappears in the power spectrum. However, as can be seen in Fig.\ \ref{FERRIMAG}(e), exemplary snapshots of this mode for $\mu_{0}M_{\mathrm{s}}=1.25\,$T demonstrate that this is not a straightforward continuation of the anti-phase breathing excitation, but instead a newly emerging non-radial mode. In detail, the top layer exhibits a mode that is reminiscent of the quadrupolar distortion discussed in Ref.\ \cite{Lin2014}, while the bottom layer still shows a pure breathing mode. Interestingly, a similar behavior is observed for high $M_{\mathrm{s}}$ values in the case of stronger interlayer exchange coupling. As shown in Fig.\ \ref{FERRIMAG}(f), the higher coupling strength implies that the bottom layer also exhibits deviations from a radially symmetric breathing mode. Comparable nonradial skyrmion eigenmodes have also been predicted by Kravchuk \textit{et al.}, albeit for the case of a single (compensated) antiferromagnetic film \cite{Kravchuk2019}, while in the present work we only observe such excitations for a sufficiently uncompensated, synthetic ferrimagnet trilayer that hosts skyrmions with relatively large diameters.  
In addition, our calculations indicate that such non-radial excitations do not occur in single ferromagnetic layers for which we have assumed identical simulation parameters as for the trilayer scenario. Therefore, the occurrence of these modes is characteristic for interlayer exchange-coupled skyrmions in uncompensated synthetic ferrimagnets. Also, it should be noted that the lowest-lying (in-phase) mode does not exhibit any deviations from the radial breathing dynamics at any of the considered values of $M_{\mathrm{s}}$.    
Furthermore, while Fig.\ \ref{FERRIMAG} only includes the case of the ac magnetic field being applied across the top layer, the same modes can be excited by exposing only the bottom layer or even the entire trilayer structure to the oscillating external field. 
Lastly, in contrast to synthetic AFMs (see Fig.\ \ref{DIFFLAYERS}), the non-vanishing total magnetization of synthetic ferrimagnets implies that the in-phase breathing mode also leads to a strong signature in the power spectrum even when the ac magnetic field is applied across all layers. In the case of synthetic AFMs, a symmetry-breaking and sufficiently large dc magnetic field along the $z$-axis is required to make the in-phase breathing mode experimentally accessible.

In conclusion, a variety of resonance peaks related to coupled breathing modes can be expected to be detected in microwave impedance spectroscopy experiments for both synthetic ferri- and antiferromagnets. In such experiments, it will be of major importance to utilize materials with damping parameters that are as low as possible in order to detect signatures of coupled breathing modes \cite{Back2020}. Ultimately, as the simulations show, the dynamic fingerprint of the coupled breathing modes, that is, the presence or absence, position, shape and number of resonances, will allow to draw conclusions about the underlying magnetic interactions and parameters. So far, the numerical calculations have implied the excitation of breathing modes by means of time-varying magnetic fields. From the experimental point of view, spin torques constitute an intriguing alternative to excite such magnetization dynamics. In the last part of this work, we will demonstrate that the excitation of coupled breathing modes in synthetic AFMs can also be realized with different types of spin torques.  

\subsection{Excitation of Breathing Modes with Spin Torques}
\begin{figure}
\centering
\includegraphics[width=7.6 cm]{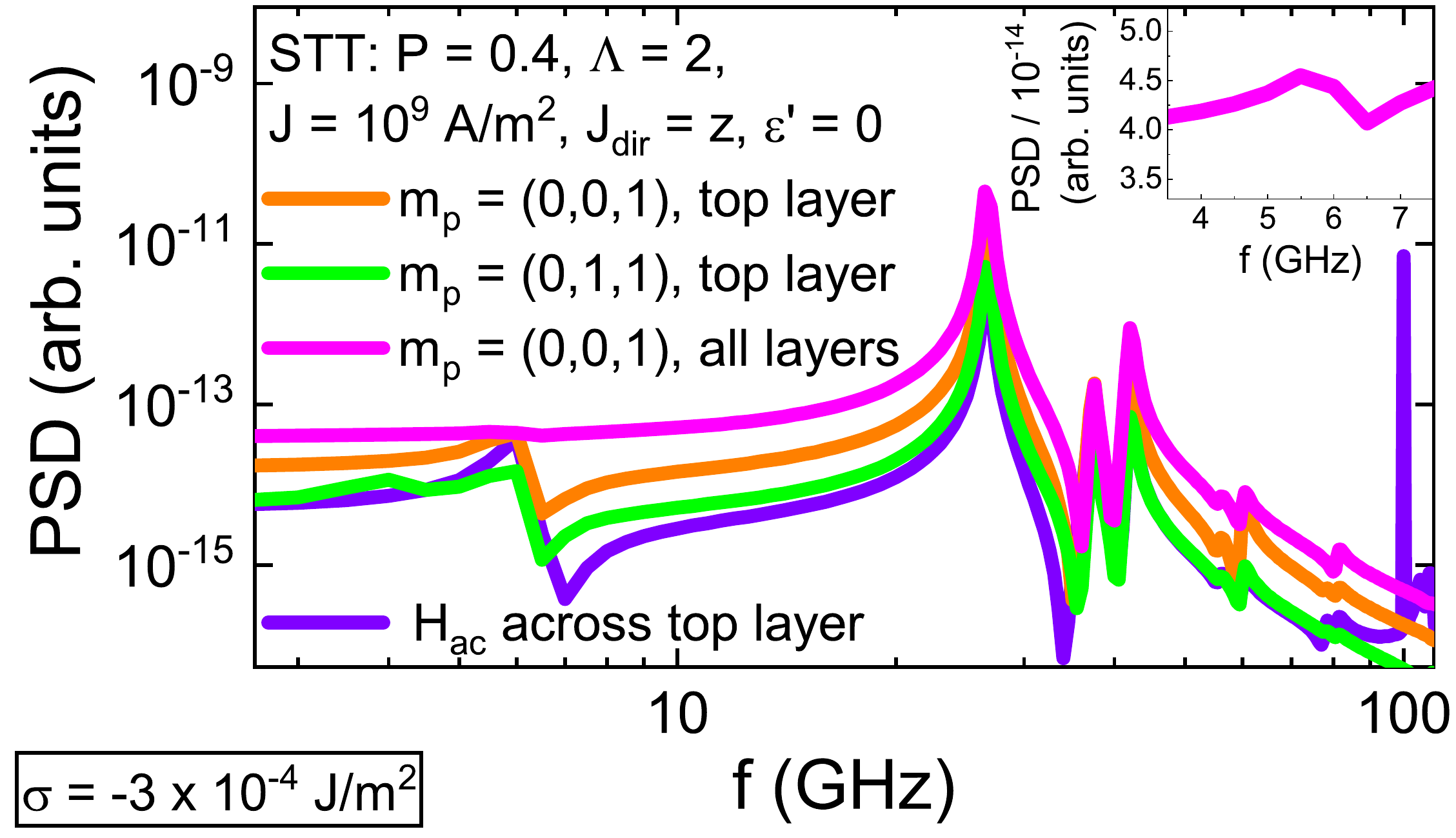}
\caption{Comparison of power spectra obtained for magnetization dynamics excited by an ac magnetic field with $f=100\,$GHz across the top layer and by a purely damping-like spin-transfer torques for an antiferromagnetic coupling strength of $\sigma = -3\times 10^{-4}\,$J/m$^{2}$. Spin currents are assumed to exhibit a spin polarization along the $(0,0,1)$ or $(0,1,1)$ direction, and to be present in either the top layer or in all layers. Inset shows an enlarged view of the pink curve from the main panel.} 
\label{STTHAC}%
\end{figure}%
In this part, we will show that the results obtained for the application of ac magnetic fields can be qualitatively reproduced when utilizing spin-transfer torques (STTs) for the excitation of the skyrmion breathing modes. In analogy to the excitation of resonance modes with ac magnetic fields, providing torque to only one ferromagnetic layer will allow to drive the system in an unbalanced manner. Moreover, it will be demonstrated that torques present in all layers can also lead to clear signatures of breathing modes in the power spectra. 

In order to model STTs in our micromagnetic simulations with \textsc{oommf}, the following term is added to the right-hand side of the LLG equation which is given in Eq.\ (\ref{eq:LLG}) \cite{Slonczewski1996, Berger1996, Xiao2004, Donahue1999}: 
\begin{equation} \label{eq:STT}
\mathrm{STT}=|\gamma_{0}|\,\beta \left[\epsilon \left( \textbf{m}\times \textbf{m}_{\mathrm{p}}\times \textbf{m}\right)-\epsilon^{\prime} \left(\textbf{m}\times \textbf{m}_{\mathrm{p}}\right) \right].
\end{equation}
For the considered synthetic AFM, we set the electron polarization direction as $\textbf{m}_{\mathrm{p}}=(0,0,1)$, that is, perpendicular to the layers, or as $\textbf{m}_{\mathrm{p}}=(0,1,1)$ in order to consider an additional in-plane component. The spin current is assumed to be injected from an additional fixed magnetic layer beneath (or on top of) the synthetic AFM structure. Furthermore, we assume that spin torques are exerted in either one or both ferromagnetic layers.  
$\epsilon$ and $\epsilon^{\prime}$ correspond to the effective spin polarization efficiency factors for the damping- and field-like torques, respectively. More specifically, $\epsilon$ can be written as
\begin{equation}
\epsilon=\frac{P\Lambda^{2}}{(\Lambda^{2}+1)+(\Lambda^{2}-1)(\textbf{m}\cdot \textbf{m}_{\mathrm{p}})}.
\end{equation}
$P$ denotes the spin polarization and $\Lambda$ is a dimensionless parameter of the model \cite{Xiao2004}.  
Finally, the other dimensionless parameter $\beta$ explicitly included in Eq.\ \ref{eq:STT} is given by  
\begin{equation}
\beta=\left|\frac{\hbar}{\mu_{0}}\right| \frac{J}{t_{\mathrm{FL}} M_{\mathrm{s}}},
\end{equation}
where $\hbar$ is the reduced Planck's constant, $\mu_{0}$ the vacuum permeability, $J$ the current density that exerts the spin-torque and $t_{\mathrm{FL}}$ the thickness of the (free) layer that is subject to the STT.  
Figure \ref{STTHAC} depicts an exemplary comparison of power spectra obtained for magnetization dynamics excited by an ac magnetic field applied across the top layer (violet curve) and by a purely damping-like STT for the case of strong antiferromagnetic coupling with $\sigma = -3\times 10^{-4}\,$J/m$^{2}$. We consider the three following scenarios for the modeled spin currents: (i) spin polarization along $(0,0,1)$ and STT only modeled in the top layer (orange curve), (ii) the extended case with STT present in both ferromagnetic layers (pink curve), and (iii) electron polarization direction along $(0,1,1)$ and STT only in the top layer (green curve). 
While some previous studies suggest that the spin polarization is strongly reduced by the spacer layer of the synthetic AFM and thus only scenarios (i) and (iii) could be regarded as realistic for the excitation of coupled breathing modes by means of STTs \cite{Zhou2020}, other works demonstrate relatively high spin-diffusion lengths $l$ for materials typically used as nonmagnetic spacers, for example $l_{\mathrm{Ru}}\approx 14\,$nm for ruthenium \cite{Eid2002}. Due to the low thickness $t_{\mathrm{NM}}=1\,$nm of the spacer layer considered in the present work, scenario (ii) may also be relevant for future experiments. In addition, we note that in our model we neglect contributions arising from interfacial effects such as the possible reflection of spin currents. Finally, as will be discussed further below, spin-orbit torques (SOTs) may constitute a promising alternative for the excitation of resonant skyrmion breathing dynamics in synthetic AFMs. In this case, assuming the presence of spin torques only in one of the ferromagnetic layers is more appropriate than the situation in scenario (ii).               
 
For the three given scenarios, all eight resonance modes occur at identical frequencies in the spectra for the simulated damping-like STT ($\epsilon \neq 0$, $\epsilon^{\prime}= 0$) with only minor differences in their magnitude compared to the modes excited by a magnetic field. While the magnitude can be controlled by the variation of parameters like $P$ or $J$, we note that the observed systematics is universal and, importantly, independent on the nature of the STT. In other words, purely field-like STTs ($\epsilon=0$, $\epsilon^{\prime}\neq 0$) or mixtures of the two STT types ($\epsilon \neq 0$, $\epsilon^{\prime}\neq 0$) lead to qualitatively similar power spectra. 
Considering the exemplary graphs in Fig.\ \ref{STTHAC}, the results for $\textbf{m}_{\mathrm{p}}=(0,0,1)$ and STTs present in one ferromagnetic layer (orange curve) show the strongest similarities with the spectrum that is related to magnetization dynamics excited by an ac magnetic field across the top layer (purple curve). However, this scenario is assessed to be challenging for current experimental realization. By contrast, the case of STTs in both magnetic layers (pink curve) can be implemented by passing a spin-polarized current through the entire synthetic AFM structure. Similar to the case of a magnetic ac field applied across all layers as shown in Fig.\ \ref{DIFFLAYERS}, the in-phase breathing mode is strongly suppressed, but still present (see inset of Fig.\ \ref{STTHAC}). 

Lastly, we will discuss the possibility to drive skyrmion breathing dynamics in synthetic AFMs by means of SOTs. In a previous work, it has been experimentally demonstrated for ferromagnetic multilayers that breathing-like excitations of skyrmions can be induced by spin-orbit torques \cite{Woo2017}. 
For the case of synthetic AFMs with perpendicular anisotropy, the exploitation of novel types of SOTs originating from materials with reduced crystalline symmetry \cite{MacNeill2016, Baek2018, Safranski2018} such as non-collinear AFMs \cite{Holanda2020, Liu2019} is desirable, since a spin polarization component along the $z$-axis is required to excite breathing modes in these systems. By contrast, a comparably high crystalline symmetry of regular spin-source materials that provide current-induced SOTs usually restricts applications to magnetic devices with in-plane anisotropy \cite{MacNeill2016}.  
For the case of spin currents generated from materials with reduced crystalline symmetries, the spin polarization can also have other contributions than solely the $z$-component. Here, by proving that breathing modes can also be excited by spin currents with $\textbf{m}_{\mathrm{p}}=(0,1,1)$ (green curve in Fig.\ \ref{STTHAC}) we conclude that experiments with novel SOTs can be expected to provide new results and possibilities with regard to dynamic excitations of skyrmions in synthetic AFMs. Note that for the example of $\textbf{m}_{\mathrm{p}}=(0,1,1)$, an additional feature in the spectrum (green curve) occurs at $f=4\,$GHz due to the simultaneous excitation of skyrmion gyration modes in this scenario.                           

\section{Summary and Conclusion}
In this work, we have numerically studied the breathing dynamics of skyrmions in synthetic AFM structures composed of two ferromagnetic layers that are separated by a nonmagnetic spacer. It was shown that varying the strength of the RKKY-like coupling through the metallic spacer layer allows for tuning the dynamic properties of in-phase and anti-phase breathing oscillations in a well-controlled way. 
In addition to that, the different response of the two major types of coupled breathing modes to alterations of magnetic parameters, such as the saturation magnetization, was presented in detail.
Moreover, the systematics of in-phase and anti-phase breathing modes was discussed for the case of synthetic ferrimagnets. Aside from the characteristic dependence of resonance frequencies on the varying saturation magnetization of the individual magnetic layers, it was demonstrated that novel, non-radial dynamic modes can emerge for a sufficiently high degree of imbalanced moments in the two ferromagnetic layers. 
Furthermore, both field- and damping-like STTs have been shown to represent an alternative means to excite skyrmion breathing dynamics in magnetic multilayers with antiferromagnetic interlayer exchange coupling. 
In conclusion, it has been proven that the spectral analysis of coupled breathing modes in synthetic AFMs offers a promising approach for the detection and detailed characterization of skyrmions. In particular, measurements of magnetoresistive signals modulated by the in-phase or anti-phase resonant breathing oscillations are expected to allow for electrical detection of magnetic skyrmions in synthetic AFMs.       

\section*{Acknowledgements}
M.\ L.\ acknowledges the financial support by the German Science Foundation (Deutsche Forschungsgemeinschaft, DFG) through the research fellowship LO 2584/1-1. This research was partially supported by the NSF through the University of Illinois at Urbana-Champaign Materials Research Science and Engineering Center DMR-1720633 and was carried out in part in the Materials Research Laboratory Central Research Facilities, University of Illinois.

\bibliography{Skyrmion_literature}


\clearpage

\end{document}